\documentclass{article}

\usepackage{arxiv}

\usepackage[utf8]{inputenc} 
\usepackage[T1]{fontenc}    
\usepackage{hyperref}       
\usepackage{url}            
\usepackage{booktabs}       
\usepackage{amsfonts}       
\usepackage{nicefrac}       
\usepackage{microtype}      
\usepackage{lipsum}

\usepackage{multicol}
\usepackage{graphicx}
\usepackage{caption}
\usepackage{subcaption}
\usepackage{multirow}
\usepackage{enumitem}
\usepackage{comment}

\newcommand{\enotesoff}{\long\gdef\enote##1{}}
\newcommand{\enoteson}{\long\gdef\enote##1{\par\noindent\fbox{\parbox {0.98\textwidth}{{\large {\bf DRAFT}  \newline} \small
\scshape ##1}}\\[0.3ex]}}
\enotesoff
\enoteson

\title{Web Mining for Estimating\\
Regulatory Blockchain Readiness}

\author{
  Elias Iosif\\
  Department of Digital Innovation, School of Business\\
  Institute For the Future (IFF)\\
  University of Nicosia\\
  Cyprus\\
  \texttt{iosif.e@unic.ac.cy}\\
   \And
  Klitos Christodoulou\\
  Department of Digital Innovation, School of Business\\
  Institute For the Future (IFF)\\
  University of Nicosia\\
  Cyprus\\
  \texttt{christodoulou.kl@unic.ac.cy}\\
  \And
  Andreas Vlachos\\
  Department of Digital Innovation, School of Business\\
  University of Nicosia\\
  Cyprus\\
  \texttt{vlachos.a@unic.ac.cy}\\
}

\begin{document}
\maketitle

\newcommand{\eg}{e.g., } 
\newcommand{\ie}{i.e., } 
\newcommand{\cf}[1]{{ cf. #1}}
\newcommand{\etal}{{et al.}\ }
\newcommand{\quotes}[1]{``#1''}

\begin{abstract}
The regulatory framework of cryptocurrencies (and, in general, blockchain tokens) is of paramount importance.
This framework drives nearly all key decisions
in the respective business areas.
In this work, a computational model is proposed
for quantitatively estimating the regulatory stance of
countries with respect to cryptocurrencies.
This is conducted via web mining utilizing web search engines.
The proposed model is experimentally validated.
In addition, unsupervised learning (clustering) is applied
for better analyzing the automatically derived estimations.
Overall, very good performance is achieved by the proposed algorithmic approach.

\end{abstract}


\section{Introduction}
\label{s:intro}
The disruptive character of blockchain technologies it is
well-acknowledged.
Indicative studies about the current challenges, as well as the future opportunities are presented in \cite{makridakis2019}, \cite{iosif2018mcis}. 
The regulatory treatment of cryptocurrencies (and, in general, blockchain tokens)
depends on national jurisdiction
\cite{drozd2017}.
In certain nations like the U.S.A, the legal treatment of cryptocurrencies varies according to several in--country jurisdictions \cite{chohan2018}.
This diversification of legal rules, highlights the importance for industry stakeholders
to assess the regulatory landscape of a certain region
prior of taking decisions regarding future business activities
(e.g., blockchain company registrations and investments).
These stakeholders may involve, but are not limited to,
individual investors, company executives, governmental authorities,
policy makers and industry professionals. 
To this date, those attempts mainly involve manual and subjective assessment of the regulatory landscape.
This assessment is a time-consuming process requiring very specific expertise.
Even if those conditions are met, 
the subjective and manual character of it may lead to erroneous estimations
affecting decision making procedures.

The present work, suggests that the proper assessment of cryptocurrency regulation
shall be based on objective information
and be able to scale into a wide range of assessed countries.
The acquisition of such information should be conducted in an algorithmic manner.
In this work, we propose an algorithmic model based on web-harvested information
for assessing cryptocurrency regulation.
This is done in order to measure the aforementioned regulatory tendency for each country.
The proposed algorithm estimates a numerical score for each country,
which represents the degree of ``non-hostile'' local cryptocurrency regulations.
The underlying methodology is based on web mining achieved through
the utilization of web search engines,
and extends metrics used in the areas of information retrieval and natural language processing.
The main contribution of this present work is that the proposed model
can suggest to policy makers which country paradigms shall be followed,
in order to stimulate local regulatory hubs within their regions.
This paradigm can be followed to assess future cryptocurrency related regulatory guidelines,
as well as any new frameworks developed to regulate exponential technologies, which are currently advancing ahead of law enforcement
\cite{fenwick2016regulation}.

\section {Related Work}
\label{s:rw}
A favourable regulatory landscape surrounding the treatment of cryptocurrencies might be able to boost the adoption by local industries and governments,
thereby attracting regional and foreign investments.
Regulation is currently the biggest obstacle to widespread cryptocurrency adoption for investment funds
\cite{cfr2021}.
Prior attempts to assess the regulatory landscape of emerging technologies have been based on surveys and assessment of legal systems, documents and articles by panels of experts.
Examples of such assessments have been implemented for areas like
autonomous vehicles by the Autonomous Vehicles Readiness Index \cite{kpmg202},
digital transformation by the Network Readiness Index
\cite{dutta19}
and
automation by the Automation Readiness Index \cite{te2008}.
To the best of our knowledge, no end--to--end algorithmic framework
was employed for the development of the
aforementioned indices.
We believe that creating an algorithmic and automated method
of assessing the legal environment of a specific sector,
will be an appropriate tool to be adopted globally for an unlimited number of countries and industries.

The same observation holds for
the specific case cryptocurrencies and, in general, blockchain tokens.
There is limited work done to date, with the aim to provide an assessment
of the legal environment surrounding Bitcoin and other cryptocurrencies.
Even though some documents have been published which provide a glimpse of the regulatory areas in some countries
(e.g., see \cite{yeoh2017},
\cite{cumming2019},
\cite{libcon2020},
\cite{gli2021})
an algorithmic model of global assessment, as mentioned before, is profoundly lacking.
Those studies do not exhibit a clear quantitative character (e.g., score-based rankings)
which could significantly reduce the assessment time.

An industry-oriented study 
has been recently published
as an initial attempt
to translate the cryptocurrency regulatory environment of 249 countries into a numerical score
\cite{ctb2020}.
This study was based on five indicators summarized as follows:
\begin{enumerate}
\item {\bf Legality of cryptocurrencies.}
This indicator examines whether cryptocurrencies are considered legal or are banned
by local governments.
The governments which have not banned the use of cryptocurrencies, are considered to be within a legal grey area which is outlined as ``dangerous''.
\item {\bf Initial Coin Offerings (ICOs) restrictions.}
The assessment of local restrictions of ICOs,
including complete bans and hostile regulations,
which could affect investors’ ability to invest and raise obstacles that developers will need to
overcome in order to develop their projects.
An example of this indicator is the assessment of the  jurisdictions in the USA,
which classifies coin offerings under securities laws,
and have forced a number of blockchain projects to reconsider the location of their company's registration. 
\item {\bf ICOs locations.}
The consideration of the number of ICOs which have officially been registered at a given nation.
\item {\bf Exchanges locations.}
The availability of exchanges in countries is important for traders and potential investors.
It is still essential for cryptocurrency users to be able to easily convert crypto--to--fiat
and fiat-to-crypto, especially when certain exchanges have faced hostile regulations in specific regions of the world.
Note that it is unclear whether this indicator assesses the number of the official registration of exchanges per country or the ability of local users per country to register for any exchange worldwide.
\item {\bf User opinions.}
This represents a subjective indicator, whereas website users are encouraged to vote for the countries they believe they have the most friendly cryptocurrency regulations. 
\end{enumerate}
The aforementioned work can be regarded as a first attempt
of providing a numerical assessment of cryptocurrency regulations worldwide.
However, it comes with a number of inefficiencies.
The procedure of obtaining the scores is highly unclear, as the formula used is not published.
Also, there is no explanation of how the score of each indicator was calculated.
Those observations imply a manual (and not fully transparent) methodology.
We believe that a proper algorithmic and automated procedure,
driven by web-harvested data
can be used and, at some extent, automate the manual approach.
The proposed algorithmic approach is formalized and experimentally validated
in the sections that follow.

\section{Model: Regulatory Stance Hypothesis}
\label{s:model}
In this section, we present a computational model
used for estimating the regulatory stance of a given country
with respect to cyptocurrencies.
That is,
the the degree of ``non-hostile'' local cryptocurrency regulations as mentioned
in Section \ref{s:intro}.
This estimation is conducted on the basis of
lexical information harvested from the world wide web
using search engines.
The regulatory stance for the countries of interest
is assumed to be reflected in the web documents indexed by the used
search engines.
Those documents are meant to thematically cover all the indicators
summarized in Section \ref{s:rw} (and, of course, more).
Next, the following aspects of the proposed model are described:
(i) the underlying hypothesis,
(ii) the model parameters
also including the needed web search queries,
and
(iii) a metric along with the respective query complexity.

{\bf Regulatory stance hypothesis based on lexical co-occurrence.}
Consider a country $c$.
The co-occurrence of positive/negative regulation-related cues with $c$'s references
within a coherent linguistic environment,
implies $c$'s tendency towards a positive/negative stance.
For example, the positive stance of a particular country with regards
to cryptocurrency regulation is expected to be observable (i.e., expressed)
in articles that are publicly available in the world wide web.
For short, the above hypothesis is also referred to as the
\emph{regulatory stance hypothesis (RSH)}.

The foundations of RSH
lie in a variant of the widely-used
\emph{distributional hypothesis of meaning (DHM)}
\cite{harris1954}
which has been applied in the area of lexical semantics
(and, in general, in the areas of natural language processing
and information retrieval)
for estimating the semantic similarity between words, as well as multi-word terms.
DHM is the core of distributional semantic models (DSMs)
suggesting that the similarity of context implies similarity of meaning \cite{iosif2015}.
A variation of DHM considers the co-occurrence of words
for deriving association measurements
which can quantify the semantic similarity
of the corresponding words, for example see \cite{bollegala2007} and \cite{iosif2009}.
One of the earliest works on word co-occurrence and associations was proposed
in \cite{church1990}.
Such semantic models are of great applicability also in the broader area
of semantic web \cite{christodoulou2015structure}.

The RSH-based model proposed in the present work constitutes a new approach.
Motivated by the hypothesis that the world wide web can be regarded
as the largest resource of lexical information,
the regulatory stance hypothesis
is implemented as a ``contrast measurement''
between positive and relative lexical cues.
Specifically, the ``contrast'' aspect is what makes the present approach different compared to previous DSM-based models
utilized for semantic similarity estimates.
Unlike the case of semantic similarity
for which the notion of ``contrast measurement'' cannot be defined,
this type of measurement is instrumental for the case of
regulatory stance.
Another difference is that
DHM and DSMs deal with pairs of words,
while the proposed RSH-based model
takes as input a single argument
(i.e., the country of interest -- see below).

Given a country $c$, the output of the model is a numerical score, $R_{c}$,
which quantifies $c$'s tendency towards a positive/negative stance.
Following the proposed RSH,
the $R_{c}$ score is computed 
according to the co-occurrence of $c$'s references with positive and negative lexical cues.
The linguistic environment where the co-occurrence is considered
is the textual context of web documents.
Assume two sets, namely, $P$ and $N$, which contain lexical cues
that imply positive and negative stance respectively.
The model parameters needed for defining $R_{c}$ are:
\begin{itemize}
\item $p_{c}$: Total number of web documents in which word $c$ co-occurs with positive cues.
\item $n_{c}$: Total number of web documents in which word $c$ co-occurs with negative cues.
\item $t_{c}$: Total number of web documents in which $c$ is mentioned.
\end{itemize}
Following the terminology of web search \cite{iosif2009},
the values above are also referred to as ``number of results'', ``number of hits''.
For a country $c$, the $R_{c}$ score is computed as follows:
\begin{equation}
\label{eq:rc}
R_{c} = \frac{p_{c}-n_{c}}{\max\{p_{c}, n_{c}\}}~\frac{p_{c}+n_{c}}{t_{c}},
\end{equation}
for $p_{c} > 0$ and/or $n_{c} > 0$.
If any of those two constraints holds, then it can be inferred that also $t_{c} > 0$.
It is observed that $R_{c}$ is expressed as the product of two factors:
(i) $\frac{p_{c}-n_{c}}{\max\{p_{c}, n_{c}\}}$,
and
(ii) $\frac{p_{c}+n_{c}}{t_{c}}$.
The first factor ranges within the $[-1, 1]$ interval
and it implements the aforementioned contrast
by taking into account the difference between $p_{c}$ and $n_{c}$.
The $\max$ operator that appears in the denominator is employed for normalization purposes.
If $p_{c} \gg n_{c}$, then $R_{c} \approx 1$,
while $R_{c} \approx -1$ when $n_{c} \gg p_{c}$. 
If $p_{c} \approx n_{c}$, then $R_{c} \approx 0$.
The normalized difference between $p_{c}$ and $n_{c}$
is weighted by the second factor
that takes values in the $(0, 1]$ interval.
The underlying idea is that the ``positive vs. negative'' signal
--as expressed by the first factor--
should be weighted analogously to number of web documents exhibiting regulation-related topics.
The sum of $p_{c}$ and $n_{c}$ in the enumerator
suggests that the polarity is irrelevant for this factor.
This sum is normalized by
the overall presence of $c$ in the world wide web
(or in a subset of the world wide web being thematically filtered
according to the proper query formulation,
as explained in the following paragraphs)
as quantified by the $t_{c}$ value.
Overall,  a positive $R_{c}$ score can be interpreted
as a tendency towards a positive stance.
Similarly, a negative $R_{c}$ score can be considered
as a tendency towards a negative stance.
Also, as $R_{c}$ approaches zero, and regardless of its sign,
the stance becomes more neutral.
Absolute neutrality can be assumed for $R_{c} = 0$,
which can take place when $p_{c} = n_{c}$
(given that $p_{c} > 0$ and $n_{c} > 0$).

In order to retrieve the values of the aforementioned model
parameters, a series of web search queries are required
as follows:
\begin{itemize}
\item $Q_{c,p}$: query(ies) for retrieving $p_{c}$
\item $Q_{c,n}$: query(ies) for retrieving $n_{c}$
\item $Q_{c,t}$: query(ies) for retrieving $t_{c}$
\end{itemize}
The above represent three query types
rather than three queries\footnote{
Having only three queries is a special case
that is explained at the end of this section.
},
i.e., for each case one or more queries can be formulated.
Each query can take the form of a text string,
which is the typical data type passed to web search APIs.
Regarding the first two query types,
$Q_{c,p}$ and $Q_{c,n}$,
the query can be the concatenation of three lexical fields.
The first field deals with the country itself, where lexical variants can be also used, e.g.,
(``USA'' $\mid$ ``United States of America'').
In this example, note the incorporation of variants
in a single query via the use of the disjunctive ``$\mid$''
operator \footnote{
Assuming that such operators are supported by the
utilized search engine.
The ``$\mid$'' operator is also referred
to as ``OR''.
}.
Let $C$ denote the set of $c$'s lexical variants.
The second field is a set of lexical entries
representing positive or negative lexical cues.
As mentioned, $P$ and $N$ denote those sets
for the case of positive and negative cues, respectively.
Each set can consist of one or more single- or multi-word entries.
Consider the following example for the case of $P$,
(``favors' $\mid$ ``strongly supports'').
In a similar way, we can have the following query fragment
for the case of $N$,
(``bans'' $\mid$ ``strictly prohibits'').
The third field can be a set of lexical entries
that can be used for thematically restricting
the overall scope of the web search.
For example, the fragment
(``cryptocurrencies'' $\mid$ ``cryptos'')
can be appended in the query.
A related example, even from a different domain,
is the ``apple fruit'' search query
that can be used for retrieving results
where the word ``apple'' is used
according to the sense of ``fruit''
(and filtering out results that may refer to the 
homonymous technology company).
Such entries are also referred to as
``pragmatic constraints''.
Let $S$ denote the set of pragmatic constraints.
Regarding the third query type, $Q_{c,t}$, a similar methodology can used.
Specifically, the query is simpler compared to the case
of $Q_{c,p}$ and $Q_{c,n}$ because the fragments that correspond to $P$ and $N$ are not needed.
Thus, for the case of $Q_{c,t}$ only the lexical entries
included in $C$ and $S$ are considered.

Based on the above definitions and discussion,
it is clear that query complexity is a key characteristic
of the proposed model.
For the case of a single country $c$, this simply
refers to the number of queries needed for estimating the
$R_{c}$ score.
There are various ways that can be followed for the formulation of queries.
This variety can be characterized by two ends as follows.
For each of the aforementioned set ($P$, $N$, $C$, and $S$),
all of its lexical entries are 
combined in a single query fragment using disjunctive operators.
According to (\ref{eq:rc}), this requires three queries per country,
while the query complexity, ${O}_{l}$,
is constant and denoted as follows
\begin{equation}
\label{eq:compl_const}
\mathcal{O}_{l}(1).
\end{equation}
Alternatively, one may formulate a separate query fragment
for each entry of the aforementioned sets.
This approach significantly increases the query complexity,
${O}_{h}$, as
\begin{equation}
\label{eq:compl_generic}
\mathcal{O}_{h}(
(\mid\!P\!\mid\mid\!C\!\mid\mid\!S\!\mid)+
(\mid\!N\!\mid\mid\!C\!\mid\mid\!S\!\mid)+
(\mid\!C\!\mid\mid\!S\!\mid)
)
\end{equation}
which can be factorized as
\begin{equation}
\label{eq:compl_generic_factor}
\mathcal{O}_{h}(
(\mid\!P\!\mid + \mid\!N\!\mid + 1)
\mid\!C\!\mid\mid\!S\!\mid
).
\end{equation}
This factorization directly reveals a cubic degree of complexity.
For the sake of clarity, it should be noted that
``$\mid\!.\!\mid$'' stands
for the cardinality of the respective set,
which is irrelevant to the
``$\mid$'' search operator also used in this paper.
It is obvious that there is a significant difference
between the two ends,
i.e., constant vs. cubic query complexity.
In general, many query formulation strategies can be devised
at any point between the two ends.
From a modeling perspective,
the aforementioned complexities represent
the lowest ($\mathcal{O}_{l}(.)$)
and the highest ($\mathcal{O}_{h}(.)$) bounds,
respectively, as far as (\ref{eq:rc}) is concerned for estimating $R_{c}$.
The advantage of employing
a complex query formulation strategy is the semantic control gained
by reducing the use of search operators like
the disjunctive ``$\mid$'' operator.
The application of the proposed model for a list of countries
exhibits linear query complexity, with respect to the size of the list,
since countries are independent.

\section{Pilot Experiment}
\label{s:exp}
In this section, we present the results of a pilot experiment.
The purpose of this pilot study is meant
as a justification test
with regards to the key proposals of this work, i.e.,
the regulatory stance hypothesis along with the computational model.
For this purpose, a validation dataset was developed consisting of $15$ countries.
The dataset was compiled by a team of three researchers being experts in the area.
The countries included the validation dataset
are, by selection, indicative examples of the positive/negative spectrum.
The experimental setup followed for this study is as follows:
\begin{itemize}
\item {\bf Web search.}
Google's Programmable Search Engine was utilized\footnote{
\url{https://developers.google.com/custom-search}
}.
Specifically, this engine was configured in order to
take into account the entire world wide web
(as indexed by Google and offered via the used service),
while the supported language was set to English.
\item {\bf Query formulation.}
Regarding $C$, only a single lexicalization was used -- the country names included in Table \ref{tab:dat}.
For the case of pragmatic constraints (members of set $S$),
the following were used:
``cryptocurrencies'', and ``bitcoin''.
Also, the \texttt{hq} parameter\footnote{
More information about the supported parameters
can be found in the documentation:
\url{https://developers.google.com/custom-search/v1/reference/rest/v1/cse/list}
}of the search engine was set
to ``central bank security exchange commission''.
In order to achieve constant query complexity,
for the case of positive and negative lexical cues
(members of $P$ and $N$)
the following cues were used:
``allows'' and ``bans'' in combination with the
``~'' search operator (i.e., ``~allows'' and ``~bans'').
This operators enables the consideration of synonyms.
\end{itemize}

The results are presented in Table \ref{tab:dat}.
Specifically, for each country of the validation dataset three scores are reported:
the final score, $R_{c}$, computed according to (\ref{eq:rc}),
as well as the values of the first and second factor of (\ref{eq:rc})
as explained in Section \ref{s:model}.

\begin{table} [!ht]
\centering
\begin{tabular}{|c|c|c|c|}
\hline
Country	($c$) & $\frac{p_{c}-n_{c}}{\max\{p_{c}, n_{c}\}}$ & $\frac{p_{c}+n_{c}}{t_{c}}$ & $R_{c}$\\
\hline
Algeria	& 0.114	& 0.305	& 0.035\\
\hline
China	& 0.673	& 0.300	& {\bf 0.202}\\
\hline
Cyprus	& 0.277	& 0.334	& 0.093\\
\hline
Egypt	& 0.313	& 0.198	& 0.062\\
\hline
France	& 0.720	& 0.180	& 0.130\\
\hline
Gibraltar & 0.544	& 0.598	& {\bf 0.325}\\
\hline
Greece	& 0.487	& 0.225	& 0.110\\
\hline
Hong Kong	& 0.497	& 0.264	& 0.131\\
\hline
Malta	& 0.386	& 0.255	& 0.098\\
\hline
Morocco	& 0.211	& 0.216	& 0.046\\
\hline
Nepal	& 0.127	& 0.247	& 0.031\\
\hline
Singapore	& 0.590	& 0.212	& 0.125\\
\hline
South Africa & 0.496 & 0.179 & 0.089\\
\hline
Switzerland	& 0.637 & 0.166	& 0.106\\
\hline
Taiwan	& 0.326	& 0.335	& 0.109\\
\hline
\end{tabular}
\caption{Pilot experiment: results.} 
\label{tab:dat}
\end{table}

It is observed that
the highest $R_{c}$ scores are obtained for
Gibraltar ($0.325$) and China ($0.202$).
Nepal and Algeria are associated with
the lowest $R_{c}$ scores, $0.031$ and $0.035$, respectively.
The role of the weighting implemented by the second factor of (\ref{eq:rc})
is clearly demonstrated for the case of France.
Without this weighting, France would be ranked as the top country according to $R_{c}$
(in contrast to the $4_{th}$ position just after Hong Kong).
The correctness of the aforementioned scores and observations
were independently justified by the three researchers.
Of course, this should be further validated by the utilization of a benchmark\footnote{
This is part of our ongoing work.
}.
Currently, and even in the absence of such a benchmark,
the three human evaluations provided sufficient indications
that the proposed model performs well (at least at the two ends of the ranked list).

\begin{table} [!ht]
\centering
\begin{tabular}{|c|c|}
\hline
Cluster ID & Countries\\
\hline
1 & China, Gibraltar \\
\hline
2 & France, Cyprus, Greece, Hong Kong, Malta, Singapore,\\
 & South Africa, Switzerland, Taiwan \\
\hline
3 & Algeria, Egypt, Morocco, Nepal\\
\hline
\end{tabular}
\caption{Pilot experiment: clustering of countries based on $R_{c}$.} 
\label{tab:clusters}
\end{table}
Another (subjective) observation is the coarse formulation of three categories taking into account the computed $R_{c}$ scores, namely,
``high $R_{c}$'', ``mid $R_{c}$'', and ``low $R_{c}$''.
In order to computationally investigate this observation,
the $k$--means clustering algorithm was applied
(with $k=3$)
over the $R_{c}$ scores listed in Table \ref{tab:dat}.
The identified clusters are shown in Table \ref{tab:clusters}.
Based on the cluster memberships, we can observe that
the two $R_{c}$ thresholds used for identifying the three clusters are 0.202 (China) and 0.062 (Egypt).
Specifically, the following
centroid values were computed and used by the clustering algorithm,
$0.264$, $0.110$, and $0.044$. 
The middle cluster (ID 2) is the most populous cluster,
which justifies the intuition that the
``high vs. mid'' and ``mid vs. low'' boundaries are blurred.
This can be further explored by applying $k$--means
(i) only over the members of the middle cluster,
and/or (ii) hierarchical clustering.
The latter does not requires the a--priory setting of $k$,
which is a desirable configuration. 
The automatically computed results of Table \ref{tab:clusters}
are in agreement with the aforementioned findings suggesting
that our model provided reasonable results
at least for the two ends
(i.e., ``high $R_{c}$'' and ``low $R_{c}$'').

\section{Conclusions}
\label{s:concl}
In this work, we have proposed a computational model
for estimating the regulatory stance of
countries with respect to cryptocurrencies.
The key idea is the regulatory stance hypothesis
suggesting that the co-occurrence of positive/negative
regulation-related cues within web documents
can be exploited for inferring the tendency towards a positive or negative stance.
The major finding of this work was the experimental justification
of the above hypothesis using web search engines.
In this framework, a metric was proposed for yielding a single score
quantifying the overall regulatory stance.
The basic operation is the ``contrast'' of search results
corresponding to positive and negative lexical cues.
A straightforward calculation of this contrast
is the normalized difference of the respective numbers of hits.
The proposed model has a series of advantageous technical characteristics
as follows:
(i) it is fully automatic saving a significant amount of time when compared
to a manual process which demands expert knowledge,
(ii) under the proper configuration, constant query complexity can be reached,
(iii) given a set of countries, full parallelization is enabled
since there are no dependencies between the countries,
i.e., one thread per country.
The only requirements are the initial query development
and the employment of a proper web search API.
Regarding the latter, the utilization of advanced search features
is recommended as they provide more expressive power to the query developer.
This can lower the query complexity (and, thus, the execution time).
Currently, we are working on the development of formal evaluation dataset
having larger country coverage compared to the present pilot study,
while we are investigating techniques from the area of automatic grammar induction (e.g., see \cite{iosif2018})
in order to automate the formulation of web search queries.
Also, the output of the proposed model fits the broader algorithmic
framework developed by authors that deals with the
robust (i.e., missing information resilience)
estimation of blockchain readiness scores \cite{iosif2021}.
One of our immediate goals is the integration of the two models.

\bibliographystyle{apalike}
\bibliography{ref}

\end{document}